\documentclass[twocolumn,
pra
,superscriptaddress
,floatfix
,aps
,10pt
,showpacs]{revtex4-2}

\usepackage{graphicx}
\usepackage{amssymb}
\usepackage{amsmath}
\usepackage{multirow}
\usepackage{array,tabularx}
\usepackage{xcolor}
\usepackage{comment}

\usepackage[colorlinks,urlcolor=blue,citecolor=blue,linkcolor=blue]{hyperref}

\usepackage{textcomp}

\usepackage{soul}
\usepackage{float}
\usepackage{upgreek}
\usepackage{makeidx}
\DeclareGraphicsExtensions{.png,.jpg,.pdf,.mps,.gif,.bmp,.eps}
\graphicspath{{figures/}}
\usepackage{physics}

\begin{document}

\title{Nonlinearity-induced reciprocity breaking in a single non-magnetic Taiji resonator}

\author{A. Mu$\tilde{\mathrm{n}}$oz de las Heras}
\email{a.munozdelasheras@unitn.it}
\affiliation{INO-CNR BEC Center and Dipartimento di Fisica, Universit$\grave{a}$ di Trento, 38123 Trento, Italy}

\author{R. Franchi}
\affiliation{Nanoscience Laboratory, Dipartimento di Fisica, Universit$\grave{a}$ di Trento, 38123 Trento, Italy}

\author{S. Biasi}
\affiliation{Nanoscience Laboratory, Dipartimento di Fisica, Universit$\grave{a}$ di Trento, 38123 Trento, Italy}

\author{M. Ghulinyan}
\affiliation{Centre for Materials and Microsystems, Fondazione Bruno Kessler, 38123 Trento, Italy}

\author{L. Pavesi}
\affiliation{Nanoscience Laboratory, Dipartimento di Fisica, Universit$\grave{a}$ di Trento, 38123 Trento, Italy}

\author{I. Carusotto}
\affiliation{INO-CNR BEC Center and Dipartimento di Fisica, Universit$\grave{a}$ di Trento, 38123 Trento, Italy}

\date{\today}

%
\begin{abstract}
We report on the demonstration of an  effective, nonlinearity-induced non-reciprocal behavior in a single non-magnetic multi-mode Taiji resonator. Non-reciprocity is achieved by a combination of an intensity-dependent refractive index and of a broken spatial reflection symmetry. Continuous wave power dependent transmission experiments show non-reciprocity and a direction-dependent optical bistability loop. These can be explained in terms of the unidirectional mode coupling that causes an asymmetric power enhancement in the resonator. The observations are quantitatively reproduced by a numerical finite-element theory and physically explained by an analytical coupled-mode theory. This nonlinear Taiji resonator has the potential of being the building block of large arrays where to study topological and/or non-Hermitian physics. This represents an important step towards the miniaturization of nonreciprocal elements for photonic integrated networks. 
\end{abstract}

\maketitle


\section{Introduction}
\label{sec:Introduction}

Lorentz reciprocity is a fundamental property of electromagnetic fields propagating in linear and non-magnetic media~\cite{Potton_2004}. In optics, it imposes that transmission through any device built by such media is independent on the direction of propagation. However, nonreciprocal elements such as optical isolators and circulators~\cite{Ribbens_1965,Jalas_2013} play a crucial role in a variety of technological applications, from unidirectional lasers to high-speed optical communications and information processing systems~\cite{Miller_2000,Kaminow_2008,Nagarajan_2010}.
The usual path to optical non-reciprocity relies on employing magnetic elements that explicitly break the time-reversal $\mathcal{T}$ symmetry~\cite{Dotsch_2005,Wang_2005,Wang_2009,Bi_2011,Shoji_2012,Ozawa2019,Yan_2020}. Due to the technical challenges associated to the monolithic integration of such elements into integrated photonics devices, alternative strategies compatible with state-of-the-art photonic technologies are being explored to circumvent Lorentz reciprocity. Time-dependent modulations of the material refractive index have been used but their operation requires an external driving field for the modulation and a large on-chip footprint~\cite{Bhandare05,yu2009complete,Fang_2012,Galland:13,doerr2014silicon}.
Since the reciprocity theorem crucially relies on the linearity of the field equations, another promising avenue is to exploit optical nonlinearities of materials.
A first configuration involved a cascade of two nonlinear resonators with different properties, which made transmission strongly direction-dependent~\cite{Fan_2012}.
Similar non-reciprocal devices were then studied exploiting complex resonator designs~\cite{Li_2020,Aleahmad_2017}, $\mathcal{P}\mathcal{T}$-symmetric coupled cavities~\cite{Peng_2014,Chang_2014}, and two-beam interactions~\cite{DelBino_2017,DelBino_2018}. Related nonlinearity-induced topology and unidirectional propagation phenomena were experimentally realized in~\cite{Maczewsky_2020}.

\begin{figure}[t]
    \centering
    \includegraphics[width=0.5\textwidth]{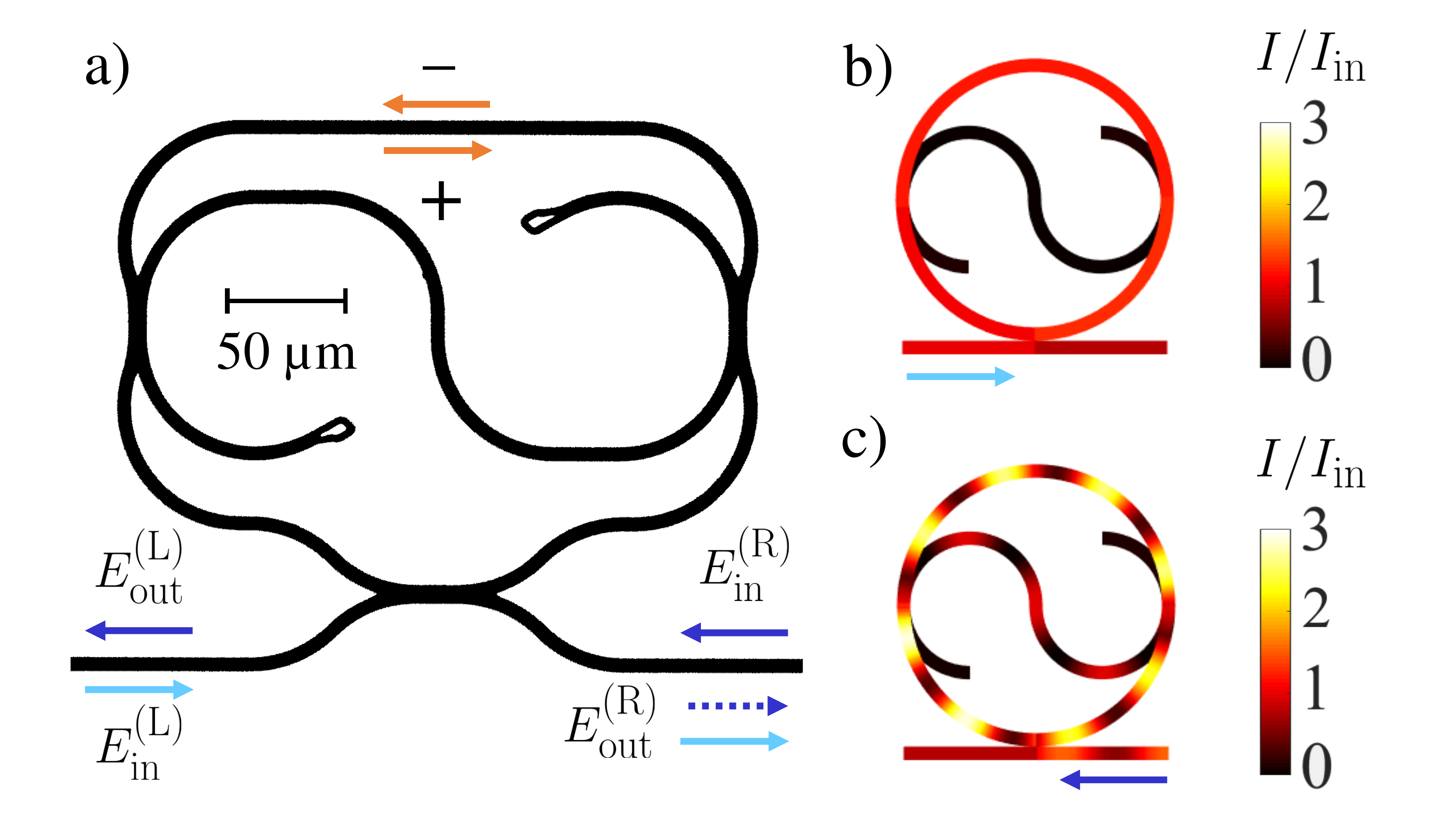}
    \caption{\textbf{a)} Optical image of the Taiji resonator (TJR) coupled to a bus waveguide. In the \textit{forward} configuration (light blue arrows) light enters the sample from the left; in the \textit{reverse} configuration (dark blue arrows) light is injected from the right. The dashed arrow represents the reflected light in the reverse case. The orange arrows indicate the propagation direction of clockwise ($+$) and counterclockwise ($-$) modes.
    \textbf{b,c)} Simulated intensity $I$ (in units of the input intensity $I_{\rm in}$) inside a TJR operating in the forward (b) and reverse (c) configurations for the same excitation frequency and power. The intensity maps are produced using the finite-element approach of Eq.~\eqref{eq:FiniteElement}. }
    \label{fig:TaijiDiagram}
\end{figure}

In this paper, we combine material nonlinearities and a broken spatial reflection symmetry to observe an effective breaking of Lorentz reciprocity in a single multi-mode Taiji resonator. Taiji micro-ring resonators (TJR) are formed by a microring with an S-shaped waveguide across as sketched in Fig.~\ref{fig:TaijiDiagram}a. In the presence of (saturable) gain, TJR have been widely exploited to obtain unidirectional operation in ring laser devices~\cite{Hohimer_1993,Kharitonov_2015,Ren_2018} and, very recently, as the basic building block of complex lattice structures acting as topological lasers~\cite{Bandres_2018}.
In the linear regime, reciprocity of the transmission across a TJR coupled to a bus waveguide is preserved and the effect of the S-shaped waveguide element is limited to an efficient unidirectional reflection~\cite{Calabrese_2020}. Still, different optical powers are accumulated in the TJR depending on the direction of illumination (Fig.~\ref{fig:TaijiDiagram}b,c). At large input powers this leads to a direction-dependent effective nonlinearity and, thus, to an effective non-reciprocal transmission. In addition to providing an intuitive explanation to the mechanism underlying this key nonreciprocal behavior, we also show the important practical advantages offered by TJR-based nonreciprocal elements in silicon photonics: a smaller footprint than previous non-magnetic proposals; an intrinsic degeneracy of the resonator modes which is protected by $\mathcal{T}$-reversal without the need for any fine tuning of the resonant frequencies of several devices; a reduced power threshold to non-reciprocity due to the use of the inter-mode nonlinearity and of the spatial overlap of the two degenerate modes.

The structure of the article is the following: In Sec.~\ref{sec:Theory} we develop an analytical coupled-mode theory of light propagation across the TJR and we make use of it to illustrate the reciprocity breaking for a TJR operating in the nonlinear regime. Sec.~\ref{sec:SamplesOpticalSetup} presents all technical details of the experimental samples and the optical setup and summarizes the finite-element model used to describe {\em ab initio} the light propagation in our device. The experimental results are shown in Sec.~\ref{sec:ExperimentalResults} and compared to the theory. Conclusions are finally drawn in Sec.~\ref{sec:Conclusions}.


\section{Theory}
\label{sec:Theory}

As it is shown in Fig.~\ref{fig:TaijiDiagram}a, a TJR supports whispering-gallery modes propagating in clockwise (CW) and counter-clockwise (CCW) directions, whose pairwise degeneracy is protected by $\mathcal{T}$-reversal. The effect of the S-shaped waveguide is to unidirectionally couple light from the CW  mode into the CCW one, while both modes display the same effective loss rate. In the \textit{forward} configuration, light accesses the bus waveguide from its left side and excites the CCW mode of the resonator through a directional coupler, but does not propagate into the S-shaped waveguide. On the other hand, in the \textit{reverse} configuration light enters the bus waveguide from the right and excites the CW mode; via the S-shaped waveguide it partially transfers excitation into the CCW one. The resulting increase of the total light intensity stored in the resonator is at the heart of the Lorentz reciprocity breaking.

A simplified model can guide our experimental analysis. Focusing on the neighborhood of a TJR resonance (resonance frequency $\omega_0$), let us assume a monochromatic excitation at $\omega$, a weak coupling of the microring to the S-shaped waveguide and to the bus waveguide (real valued coupling coefficients $k_{\rm S}, k_{\rm w} \ll 1$), and a power dependent refractive index of the TJR $n=n_{\rm L}+n_{\rm NL} I$, where $I$ is the optical intensity in the TJR (Fig.~\ref{fig:TaijiDiagram}b,c) and $n_{\rm L}$, $n_{\rm NL}$ are the linear refractive index and the nonlinear coefficient. We can adopt the widely used temporal coupled-mode theory~\cite{Walls_2012,Mher_Fano} and write the steady state electric field amplitudes $E_{\pm}$ in the CW ($+$) and CCW ($-$) modes as 
%
\begin{align}
\label{eq:CoupledMode1}
    \omega E_+&
    =\omega_{0}E_{+}-\frac{n_{\rm NL}}{n_{\rm L}}\omega_{0}(|E_{+}|^2+g|E_{-}|^2)E_{+}-i\gamma_{\rm T} E_{+}\nonumber\\
    &-\frac{c}{Ln_{\rm L}}k_{\rm w}E^{\rm(R)}_{\rm in},\\
    \omega E_-&
    =\omega_{0}E_{-}-\frac{n_{\rm NL}}{n_{\rm L}}\omega_{0}(|E_{-}|^2+g|E_{+}|^2)E_{-}-i\gamma_{\rm T} E_{-}\nonumber\\
    &-\frac{c}{Ln_{\rm L}}k_{\rm w}E^{\rm(L)}_{\rm in}-i\frac{c}{L n_{\rm L}}2k^{2}_{\rm S}e^{i\frac{\omega}{c} n_{\rm L}L_{\rm S}}E_{+}\,. \label{eq:CoupledMode2}
\end{align}
Here, $L$ and $L_{\rm S}$ are the microring and the S-shaped waveguide lengths, $c$ is the vacuum speed of light, and $\gamma_{\rm T}$ is the TJR loss rate (see later). $E^{\rm (L)}_{\rm in}$ and $E^{\rm (R)}_{\rm in}$ are the input fields exciting the ring from the left and the right of the bus waveguide, while the output fields to the left and to the right are $E^{\rm (L)}_{\rm out}=\sqrt{1-k^{2}_{\rm w}}\,E^{(R)}_{\rm in}+ik_{\rm w}E_{+}$ and $E^{\rm (R)}_{\rm out}=\sqrt{1-k^{2}_{\rm w}}\,E^{(L)}_{\rm in}+ik_{\rm w}E_{-}$, respectively.

The strength of the nonlinearity is quantified by $n_{\rm NL}=n_{\rm K}+n_{\rm T}$ which is the sum of the Kerr $n_{\rm K}$ and thermal $n_{\rm T}$ nonlinearities. Of particular interest is the parameter $g$ which describes the nature of the TJR nonlinearity. $g=2$ represents a spatially local Kerr nonlinearity~\cite{Chiao_1966,Kaplan_1983,Ghalanos_2020}, while $g=1$ describes a thermo-optic nonlinearity~\cite{Ilchenko_1992}. Note that the thermo-optic nonlinearity is mediated by a homogeneous heating of the TJR depending on the total energy that is dissipated in it. As shown in Appendix~\ref{sec:App_CM}, intermediate situations in which both processes contribute to $n_{\rm NL}$ are described by $g=(2n_{\rm K}+n_{\rm T})/(n_{\rm K}+n_{\rm T})$. 
In our specific silicon-based TJR, we have $n_{\rm T} \gg n_{\rm K}$ and thus $g\simeq 1$. 

\begin{figure}[t]
    \centering
    \includegraphics[width=0.5\textwidth]{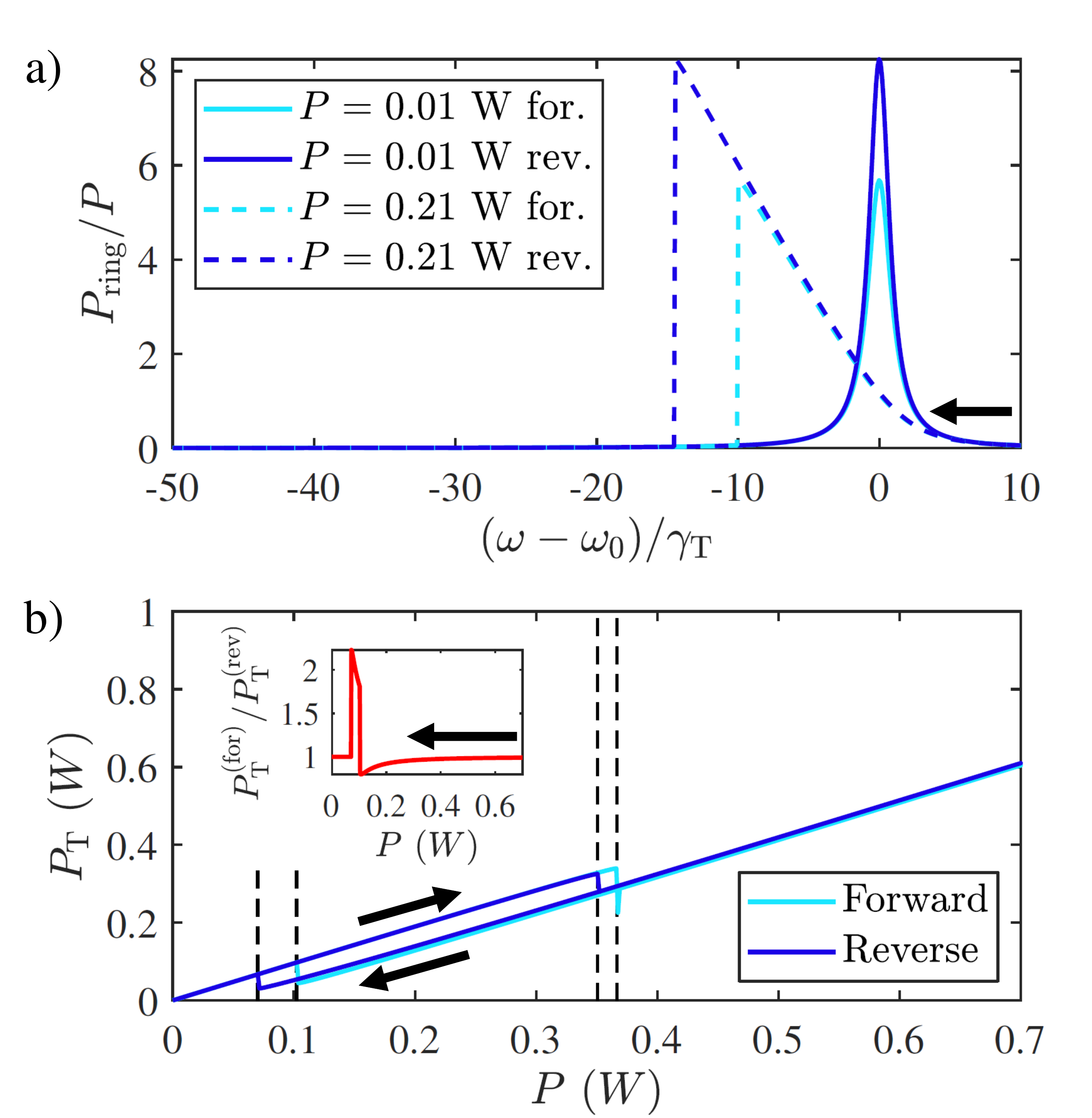}
    \caption{\textbf{a)} Simulated internal power $P_{\rm ring}$ for a downwards frequency sweep (black arrow) at two input powers $P$. Light (dark) blue curves refer to the forward (reverse) configuration. \textbf{b)} Simulated transmitted power $P_{\rm T}$ for a fixed frequency $\omega-\omega_{0}\simeq -4\gamma_{\rm T}$ in the forward and reverse configurations. Black arrows indicate the sweep direction. Black dashed lines label the threshold powers. The inset shows the ratio $P_{\rm T}^{\rm (for)} / P_{\rm T}^{\rm (rev)}$ of the transmitted powers in the forward and reverse configurations along a decreasing $P$ ramp. Parameters for both panels: $L \simeq 850$ $\mu$m, $L_{\rm S}=L/2$, $k_{\rm w}=k_{\rm S}=0.14$, $\gamma_{\rm A}=5.7 \times 10^{9}$ s$^{-1}$, $n_{\rm L}=1.83$, $n_{\rm T}=8.8\times 10^{-13}$ cm$^2$/W, $A=0.66$ $\mu$m$^2$, and $g=1$.}
    \label{fig:NewFigure3}
\end{figure}
\begin{figure*}[t]
    \centering
    \includegraphics[width=\textwidth]{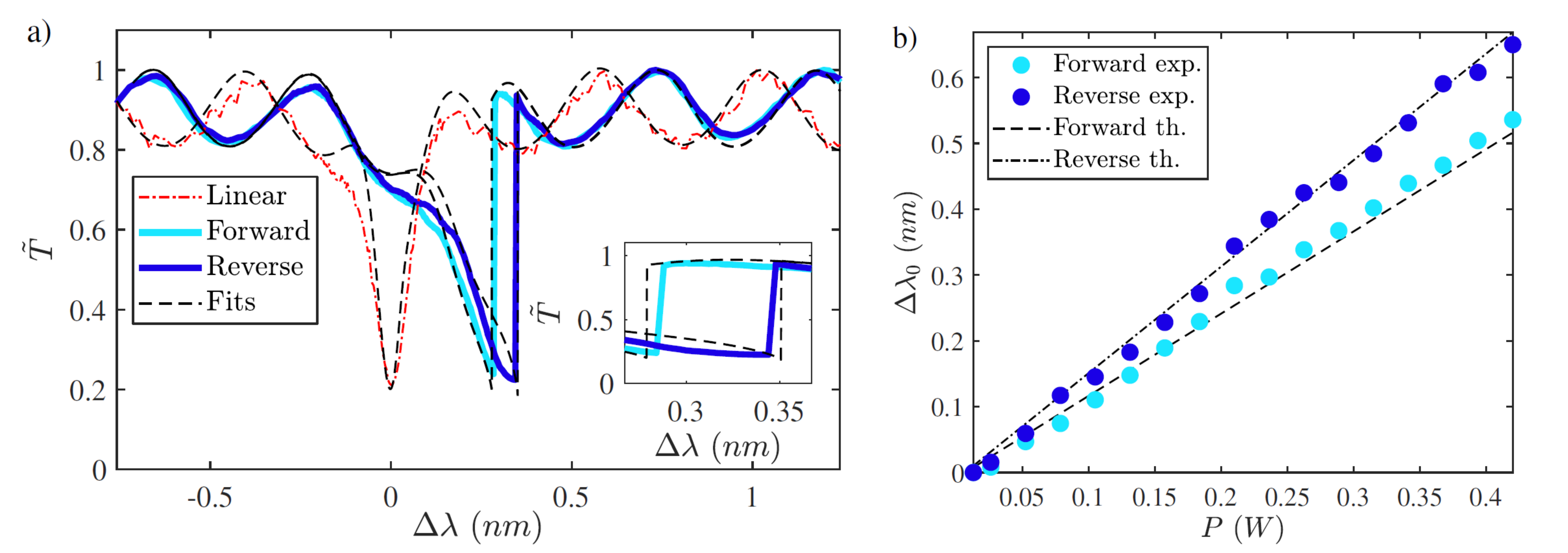}
    \caption{\textbf{a)} 
    Normalized transmittance $\tilde{T}=T/T_{\rm max}$ as a function of the detuning  $\Delta\lambda=\lambda-\lambda_{0}$ from the linear resonance wavelength for an upwards wavelength sweep  (black arrow). The red dotted line represents the experimental transmittance in the linear regime for an input power $P \simeq 0.01$ W. The light (dark) blue solid curve corresponds to the experimental transmittance in forward (reverse) configuration at a fixed input power $P=0.21$ W. The black dashed lines are theoretical fits of each experimental curve produced using Eq.~\eqref{eq:FiniteElement}. The inset gives a magnified view of the region of strongest nonreciprocal behavior. \textbf{b)} Shift $\Delta\lambda_{0}$ of the resonance wavelength as a function of the input power $P$ for the resonance displayed in panel a). Light (dark) blue circles correspond to experimental data in forward (reverse) configuration. The black dashed (dotted) line is a linear fit to the theoretically calculated $\Delta\lambda_{0}$ in forward (reverse) configuration.}
    \label{fig:ReciprocityBreaking}
\end{figure*}

The total effective loss rate $\gamma_{\rm T}=\gamma_{\rm A}+ck^2_{\rm w}/(2Ln_{\rm L})+ck^2_{\rm S}/(Ln_{\rm L})$ is the same for the CW and CCW modes. It results from the sum of absorption $\gamma_{\rm A}$ and radiative losses into the bus and the S-shaped waveguides. Yet, the last term of \eqref{eq:CoupledMode2} shows the asymmetric mode coupling introduced by the S-shaped waveguide: while all light lost by the CCW mode into the S-shaped waveguide is radiated away, part of the light transferred by the CW mode into the S-shaped waveguide is re-injected in the CCW mode. This is illustrated in Fig.~\ref{fig:TaijiDiagram}b,c: In the forward configuration, only the CCW mode is excited and no reflection occurs. In the reverse one, both modes are excited, and marked interference fringes are visible in the intensity profile. As a result, for the same frequency and input power, the unidirectional mode coupling leads to a larger intensity inside the TJR in the reverse configuration than in the forward one.

While this asymmetry does not affect reciprocity in the linear regime of weak excitations~\cite{Calabrese_2020}, it has a major impact on the nonlinear response to strong fields. In Fig.~\ref{fig:NewFigure3}a, we show the numerical prediction of the coupled-mode equations (\ref{eq:CoupledMode1}-\ref{eq:CoupledMode2}) for the internal power propagating in the ring $P_{\rm ring} \simeq A\epsilon_{0}c n_{\rm L}(|E_{+}|^2+|E_{-}|^2)/2$ (with $A$ the waveguide cross section and $\epsilon_{0}$ the vacuum permittivity). The input power $P\simeq A\epsilon_{0}c n_{\rm L}|E_{\rm in}|^2/2$ is kept at constant values while the frequency is scanned downwards across a resonance in forward and reverse configurations.
At low input power the usual Lorentzian peak is found; though, because of the S-shaped waveguide, the internal power $P_{\rm ring}$ is higher in the reverse configuration.
For a larger input power, the nonlinear refractive index causes a shift of the resonance proportional to the TJR internal power. In our $n_{\rm NL}>0$ case, the shift is towards lower frequencies. At sufficiently large powers, both curves display a sudden downward jump right after the resonance, as typical in optical bistability~\cite{Boyd,Butcher,Gibbs_1985,RamiroManzano_2013,Bernard_2017}. Note that the position of the jump depends on the direction: the larger internal intensity in the reverse configuration allows for a larger shift of the resonance before jumping. This difference is responsible for a frequency window where the internal intensity in the two configurations is strikingly different.

This key feature is an example of the nonlinear breaking of reciprocity and is further illustrated in Fig.~\ref{fig:NewFigure3}b. Here, we show the simulated transmitted power $P_{\rm T}$ as a function of the input power $P$ for a fixed incident frequency $\omega-\omega_{0}\simeq -4\gamma_{\rm T}$ in the optical bistability regime. As the input power grows, $P_{\rm T}$ displays a linear increase up to a threshold $P_1$ where a sudden jump down onto another stable solution occurs. On the way back, for decreasing $P$, the threshold $P_2$ for the upwards jump is such that $P_2 \ll P_1$. Thus, a bistability loop opens that can be understood as a metastability in a first order phase transition. Once again, the different values of the internal intensity $P_{\rm ring}$ in the forward and reverse cases lead to markedly different values for $P_1$ and $P_2$. This results in intensity windows where the transmitted powers in the two directions are very different, as shown in the inset. 


\section{Samples and optical setup}
\label{sec:SamplesOpticalSetup}

In order to verify our predictions, we fabricated integrated TJR devices using silicon oxynitride (SiON) single mode channel waveguides as shown in Fig.~\ref{fig:TaijiDiagram}a ~\cite{Calabrese_2020}. The waveguide cross section is $A=0.66$ $\mu$m$^2$. As measured in~\cite{Trenti_2018} at a wavelength $\lambda \simeq 1550$ nm, $n_{\rm L}=1.83$ and the linear absorption coefficient $\alpha=0.35\times10^{-4}$ $\mu$m$^{-1}$. We estimated $n_{\rm T}= (8.8 \pm 0.4)\times 10^{-13}$ cm$^2/$W from the fit of our experimental data, much larger than the typical Kerr nonlinear refractive index $n_{\rm K}=8 \times 10^{-16}$ cm$^{2}$/W of the material \cite{Trenti_2018}. The length of the bus waveguide is approximately the same on both sides of the TJR. In order to avoid undesired mode couplings, the ends of the S-shaped waveguide have been designed with a particular geometry (looking as rhomboids in Fig \ref{fig:TaijiDiagram}a) which prevents back-reflections~\cite{Castellan_2016}. 

The optical setup employs a fiber-coupled continuous wave tunable laser operating at the wavelength range spanning from $1490$~nm to $1640$~nm. Its output is coupled to an Erbium-Doped Fiber Amplifier to get high power. Then, a polarization control stage sets the input light to the TM (transverse magnetic) polarization. Light is coupled in the bus waveguide by butt-coupling through a tapered fiber. The transmitted light is collected by a fiber and sent to an InGaAs detector. In order to switch between the forward and the reverse configurations, the sample is simply turned without any other change in the setup.

As it was reported in~\cite{Calabrese_2020}, the response of our samples is made more complicated by Fabry-Pérot oscillations in the transmittance due to reflections at the bus waveguide facets. Even though this does not affect the qualitative predictions of the coupled mode theory, a quantitative description requires taking this effect into account as well as relaxing other implicit approximations. To this purpose, we have built a more refined theory based on the solution of the nonlinear Helmholtz equation in our specific geometry. As it is detailed in Appendix~\ref{sec:App_FE} and in~\cite{Calabrese_2020}, the ring resonator as well as the S-shaped and the bus waveguides are appropriately segmented  and propagation of the forward- and reverse-propagating waves along each segment of length $\Delta z \gg \lambda$ is described by the steady-state condition
\begin{eqnarray}
\label{eq:FiniteElement}
&&E_{\pm}(z \pm \Delta z)=\exp\Big\{i\Big[\left(n_{\rm L}+i\alpha\frac{c}{\omega}\right) \\
&&+n_{\rm T}\frac{\Delta z}{L}\sum^{N}_{j=1}\left(|E_{\pm}(z_{j})|^2+|E_{\mp}(z_{j})|^2\right)\Big]\frac{\omega}{c}\Delta z\Big\}E_{\pm}(z)\,. \nonumber
\end{eqnarray}
where the thermo-optic nonlinear refractive index of each element results from the averaged power within it. Mixing of the field in the different elements is provided by directional couplers, while reflection at both ends of the bus waveguide is taken as lossless with reflection amplitude $ik_{\rm m}$. As discussed in Appendix~\ref{sec:App_ComparisonCM-FE}, this finite-element theory matches the coupled-mode theory in the appropriate limits and can be used to quantitatively fit our experimental data.





\section{Experimental results}
\label{sec:ExperimentalResults}

To probe the TJR response, we performed sweeps of the laser wavelength at fixed powers around the cold TJR resonance at $\lambda_{0}=1545.76$~nm. When operated at a small input power, $P \lesssim 0.03$ W, the thermal nonlinearity plays no role and the cold sample behaves as a linear device. Indeed, we observed the same transmittance $T$ when pumped from the left or from the right. The linear transmittance normalized to its maximum value $\tilde{T}=T/T_{\rm max}$ is shown as a function of the detuning  $\Delta\lambda=\lambda-\lambda_{0}$ from the resonance wavelength as the red dash-dotted line of Fig.~\ref{fig:ReciprocityBreaking}a. From a fit of this curve with the theoretical model (black dashed lines), we extracted the coefficients $k_{\rm w}=0.49 \pm 0.02$, $k_{\rm S}=0.14 \pm 0.03$ and $k_{\rm m}=0.24 \pm 0.04$. A list of all parameters employed in the fits of the experimental data can be found in Appendix~\ref{sec:App_FitParameters}.

\begin{figure}[t]
    \centering
    \includegraphics[width=0.5\textwidth]{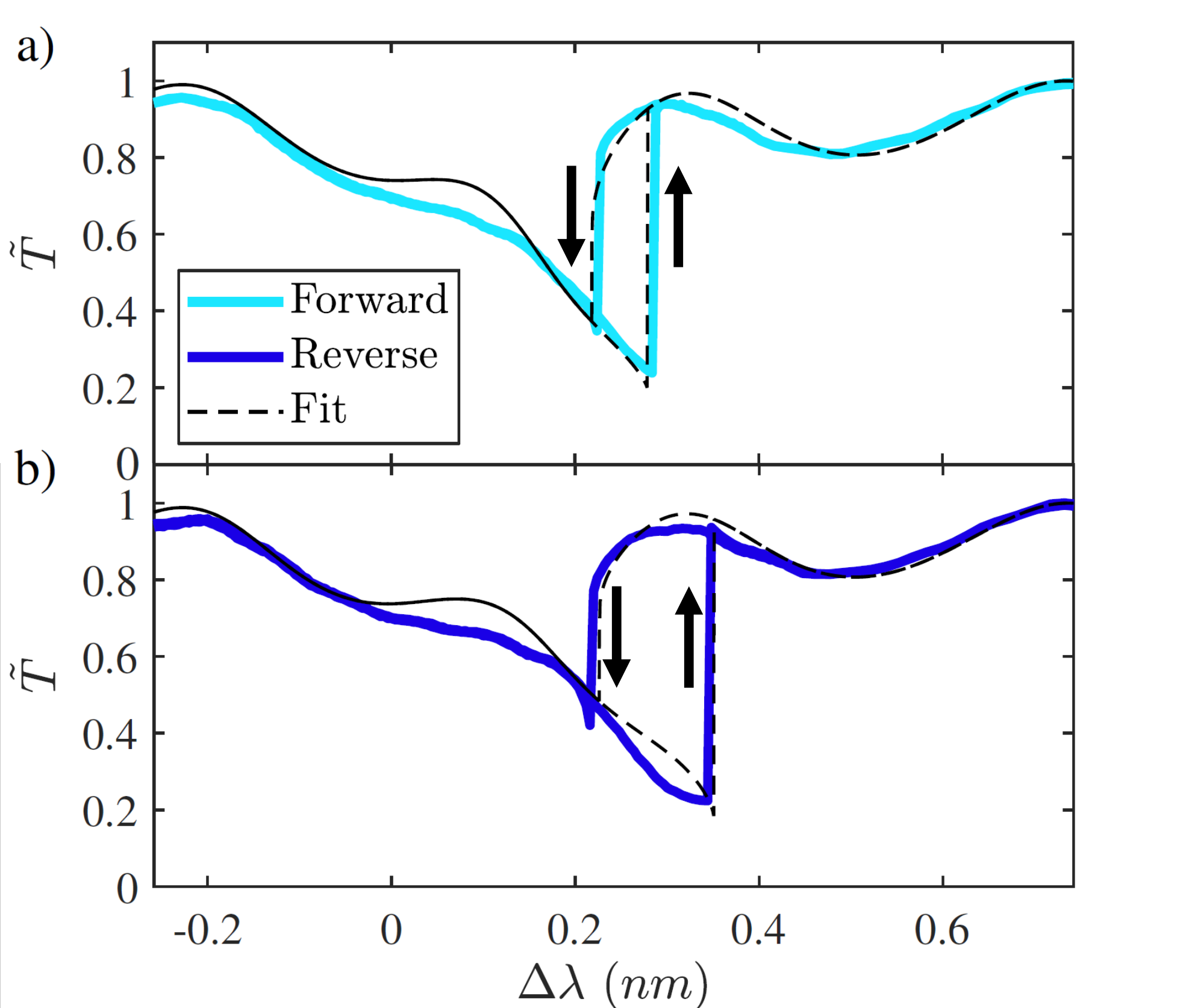}
    \caption{Optical bistability in the normalized transmittance $\tilde{T}=T/T_{\rm max}$ along a round-trip sweep of the laser detuning $\Delta\lambda$ across the resonance at a fixed input power $P=0.21$ W. The ramp direction is indicated by the black arrows. Experimental data are shown as solid lines while the black dashed lines are the fits obtained with the finite-element theory of Eq.~\eqref{eq:FiniteElement}. \textbf{a)} Forward configuration (light blue). \textbf{b)} Reverse configuration (dark blue).}
    \label{fig:OpticalBistability}
\end{figure}

In order to investigate the nonlinear response, we then performed upwards wavelength ramps at larger input powers (an example for $P=0.21$~W is shown in Fig.~\ref{fig:ReciprocityBreaking}a). As predicted by the model (Fig.~\ref{fig:NewFigure3}), the resonance dip in $\tilde{T}$ is pushed towards longer wavelengths by the nonlinearity due to the accumulated power in the TJR up to the threshold where $\tilde{T}$ jumps back to a value close to one. In the reverse configuration, since a larger intensity is present inside the resonator for the same input power, the resonance wavelength experiences a larger displacement than in the forward configuration. This is our first evidence of a nonlinearity-induced nonreciprocal behavior.
A magnified view around the transition points where the difference is the largest is shown in the inset of Fig.~\ref{fig:ReciprocityBreaking}a. Here, the ratio of the transmittance in the two directions reaches a value $T_{\rm for}/T_{\rm rev} \simeq $ 6 dB at  $\Delta\lambda\simeq 0.34$~nm. 
The displacement $\Delta\lambda_{0}$ of the resonance from its linear position as a function of the input power $P$ is displayed in Fig.~\ref{fig:ReciprocityBreaking}b: 
the non-reciprocity is clearly visible as a larger slope in the reverse case because of the higher internal power. The small deviations of the experimental points (circles) from a linear fit of the theory (lines) are mostly due to the Fabry-P\'erot oscillations. More details on the difference between the resonance shift shown in Fig.~\ref{fig:ReciprocityBreaking}b and that produced by a Kerr nonlinearity of the same magnitude can be found  in the Appendix~\ref{sec:App_ShiftThermalKerr}.

Further insight in this physics is provided by the bistability loops that are observed when comparing the transmittance along frequency ramps in upwards and downwards directions at the same fixed input power. In agreement with the coupled-mode theory of Fig.~\ref{fig:NewFigure3}, the experimental observations in Fig.~\ref{fig:OpticalBistability} show that the bistable loop observed in the reverse configuration (bottom) is wider than the one in the forward configuration (top), due to the different feedback caused by the S-shaped waveguide in the TJR. This is also reproduced by the finite element model (dashed lines).


\section{Conclusions}
\label{sec:Conclusions}

We have demonstrated how the combination of optical nonlinearities and a spatially asymmetric design gives rise to an effective violation of reciprocity in a single non-magnetic multi-mode Taiji resonator. This device goes one step beyond previous multi-resonator proposals and realizations in terms of integrability in silicon photonics circuits. The simplicity of its design allows a transparent theoretical analysis and facilitates its use as the unit cell of more complex structures. 

The effective breaking of reciprocity is visible in the dependence of both the nonlinear shift of the resonance wavelength and the width of the optical bistability loop on the direction of illumination. The experimental observations are quantitatively reproduced by the theory and the effect is intuitively understood by a coupled-mode theory in terms of the asymmetric coupling introduced by the S-shaped waveguide and the consequently different strength of the nonlinear feedback effect.

Future work will address the technologically important issue of optimizing the nonreciprocity by employing Taiji resonators built of highly nonlinear materials in a properly designed critical coupling regime. The nonreciprocal bandwidth can also be extended by engineering an optimized coupling with the S-shaped waveguide allowing to achieve simultaneously a high quality factor and a larger exchange of energy between the counterpropagating modes. 
Further prospective research lines will include building atop~\cite{Shi_2015} to investigate the limitations set by dynamical reciprocity on the actual performance of our TJR device as a nonlinear optical isolator; and exploring the richer variety of nonlinearity-induced nonreciprocal effects in large arrays of resonators where they interplay with non-trivial band topologies and topological edge states~\cite{Ozawa2019,Smirnova_2020}. On the long run, this will be ultimately extended to non-Hermitian systems with gain and losses \cite{Parto_2021}.


\section{Acknowledgements}

We acknowledge financial support from the Provincia Autonoma di Trento and the Q@TN initiative. S.B. acknowledges funding from the MIUR under the project PRIN PELM (20177 PSCKT). The authors would like to thank F. Ramiro Manzano, A. Calabrese and H. M. Price for continuous and insightful exchanges and E. Moser for technical support.


\appendix


\section{General coupled-mode theory}
\label{sec:App_CM}

The Taiji resonator (TJR) coupled-mode theory steady-state equations in the most general case featuring a combination of local and nonlocal nonlinearities read
\begin{align}
    \omega E_+
    &=\omega_{0}E_{+}-\frac{n_{\rm K}}{n_{\rm L}}\omega_{0}(|E_{+}|^2+2|E_{-}|^2)E_{+}\nonumber\\
    &-\frac{n_{\rm T}}{n_{\rm L}}\omega_{0}(|E_{+}|^2+|E_{-}|^2)E_{+}-i\gamma_{\rm T} E_{+}
    -\frac{c}{Ln_{\rm L}}k_{\rm w}E^{\rm(R)}_{\rm in},\\
    \omega E_-
    &=\omega_{0}E_{-}-\frac{n_{\rm K}}{n_{\rm L}}\omega_{0}(|E_{-}|^2+2|E_{+}|^2)E_{-}\nonumber\\
    &-\frac{n_{\rm T}}{n_{\rm L}}\omega_{0}(|E_{-}|^2+|E_{+}|^2)E_{-}-i\gamma_{\rm T} E_{-}
    -\frac{c}{Ln_{\rm L}}k_{\rm w}E^{\rm(L)}_{\rm in}\nonumber\\
    &-i\frac{c}{L n_{\rm L}}2k^{2}_{\rm S}e^{i\frac{\omega}{c} n_{\rm L}L_{\rm S}}E_{+} .
\end{align}
By properly grouping both nonlinear terms it is easy to show that
\begin{align}
    \omega E_+
    &=\omega_{0}E_{+}-\frac{n_{\rm K}+n_{\rm T}}{n_{\rm L}}\omega_{0}\left(|E_{+}|^2+\frac{2n_{\rm K}+n_{\rm T}}{n_{\rm K}+n_{\rm T}}|E_{-}|^2\right)E_{+}\nonumber\\
    &-i\gamma_{\rm T} E_{+}
    -\frac{c}{Ln_{\rm L}}k_{\rm w}E^{\rm(R)}_{\rm in},\\
    \omega E_-
    &=\omega_{0}E_{-}-\frac{n_{\rm K}+n_{\rm T}}{n_{\rm L}}\omega_{0}\left(|E_{-}|^2+\frac{2n_{\rm K}+n_{\rm T}}{n_{\rm K}+n_{\rm T}}|E_{+}|^2\right)E_{-}\nonumber\\
    &-i\gamma_{\rm T} E_{-}
    -\frac{c}{Ln_{\rm L}}k_{\rm w}E^{\rm(L)}_{\rm in}-i\frac{c}{L n_{\rm L}}2k^{2}_{\rm S}e^{i\frac{\omega}{c} n_{\rm L}L_{\rm S}}E_{+} ,
\end{align}
and therefore
\begin{align}
    n_{\rm NL}=n_{\rm K}+n_{\rm T} ,
    \;\;\;\;
    g=\frac{2n_{\rm K}+n_{\rm T}}{n_{\rm K}+n_{\rm T}} .
\end{align}
Upon replacing $\omega \rightarrow i\partial_t$, these equations can be used to describe the evolution of the system in time under the assumption of a temporally local nonlinearity. While this is usually a good approximation for Kerr media, one must keep in mind that thermal nonlinearities are typically slow so this approximation must be explicitly verified on a case-by-case basis.


\section{Finite-element model}
\label{sec:App_FE}

\begin{figure}[t]
    \centering
    \includegraphics[width=0.5\textwidth]{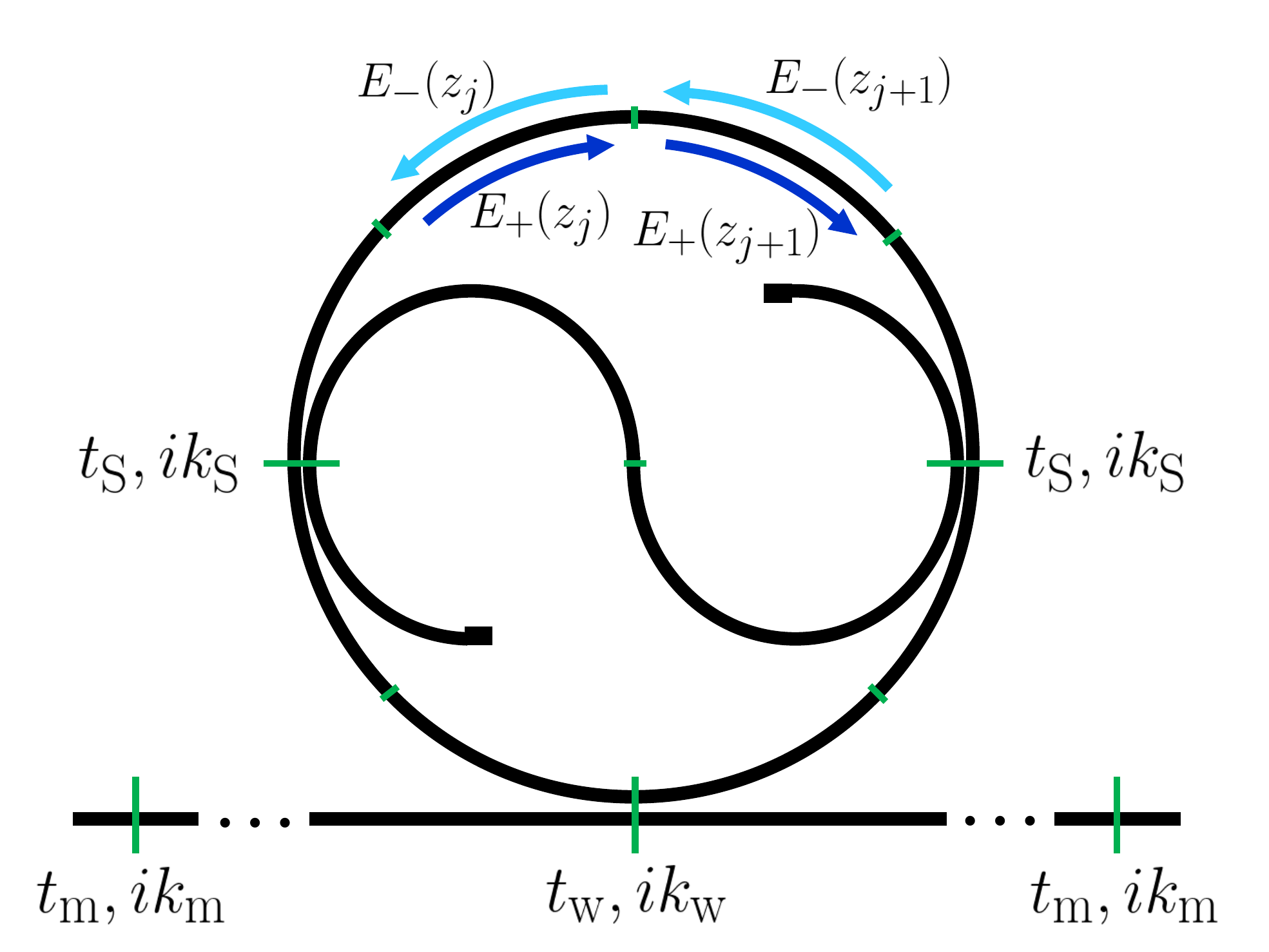}
    \caption{Scheme of the modelled Taiji resonator (TJR) showing the electric field amplitudes $E$ inside each segment $z_{j}$ (delimited by the short transverse lines) and the transmission and coupling amplitudes at each directional coupler (represented by the long transverse lines) $t_{\rm w,S,m}$ and $ik_{\rm w,S,m}$.}
    \label{fig:ModelTaiji}
\end{figure}

In order to derive the finite-element equations to be employed in the fits of the experimental results
we divided each waveguide in the sample into segments of equal length $\Delta z_{\rm r,S,w} = L_{\rm r,S,w}/N_{r,S,w} = z_{j}-z_{j-1}$, being $L_{\rm r,S,w}$\footnote{Note that here and in the following we use the compact notation of repeated indices for different quantities, i.e. $L_{\rm r,S,w}$ corresponds to $L_{\rm r}$, $L_{\rm S}$ and $L_{\rm w}$.} the total length of each component (ring, S, and bus waveguides), $N_{\rm r,S,w}$ the number of segments in which it is divided, and $z_{j}$ the spatial coordinate of each segment $j=1,...,N_{\rm r,S,w}$. The segment length in each case is assumed to be much longer than the light wavelength, i.e. $\Delta z_{\rm r,S,w}\gg\lambda$. The aim of the finite-element model is to relate the amplitude of the electric field $E$ at a position $z_{j+1}$ to the amplitude in the precedent segment $z_{j}$. To simplify the notation, on the following we will drop the subindices referring to the sample components as the equations are valid regardless of which waveguide is considered. We started from a modified Helmholtz's equation including a local Kerr nonlinearity with refractive index $n_{\rm K}$, and a nonlocal thermo-optical nonlinearity with refractive index $n_{\rm T}$. It reads
\begin{align}
\frac{\partial^2 E}{\partial z^2}&=-\left(\frac{\omega}{c}\right)^2\Bigg[\left(n_{\rm L}+i\alpha\frac{c}{\omega}\right)+n_{\rm K}|E(z)|^2\nonumber\\
&+n_{\rm T}\frac{\Delta z}{L}\sum^{N}_{j=1}|E(z_{j})|^2\Bigg]^2 E(z),
\label{eq:HelmholtzEq}
\end{align}
where $\omega$ and $c$ correspond to the angular frequency and speed of light in free space, respectively, $n_{\rm L}$ is the linear refractive index, and $\alpha$ is the absorption coefficient. Note that the thermal nonlinearity shifts the refractive index at each sample component proportionally to the average intensity inside it. In our case we have that $n_{\rm L} \gg n_{\rm K,T}|E(z)|^2$ $\forall z$, which implies that the field oscillation due to the linear part of the material's response will be much faster than that of the nonlinear part. Therefore one can employ the Ansatz
\begin{align}
E(z)&=E_{+}(z)+E_{-}(z)\nonumber\\
&=\xi_{+}(z)e^{i\frac{\omega}{c}(n_{\rm L}+i\alpha c/\omega) z}+\xi_{-}(z)e^{-i\frac{\omega}{c}(n_{\rm L}+i\alpha c/\omega) z} ,
\label{eq:HelmholtzAnsatz}
\end{align}
where $E_{\pm}$ are the electric field amplitudes and $\xi_{\pm}$ are the slowly-evolving parts of the field propagating in the clockwise ($+$) and counterclockwise ($-$) directions.

After inserting Eq.~\eqref{eq:HelmholtzAnsatz} into Eq.~\eqref{eq:HelmholtzEq} we use the rotating wave approximation to neglect those terms oscillating with spatial frequency on the order of $\omega/c$ or faster, which average to zero in a segment much longer than the optical wavevelength, as well as the smaller terms proportional to $\partial^2\xi_{+,-}/\partial z^2$ and those of order $\mathcal{O}(n^2_{\rm K,T})$. Identifying the energy-conserving processes we obtain
\begin{align}
\label{eq:FiniteElementAppendix1}
\frac{\partial \xi_{\pm}}{\partial z}&=\pm i\frac{\omega}{c}\Bigg[n_{\rm K}\left(|\xi_{\pm}(z)|^2+2|\xi_{\mp}(z)|^2\right)
\nonumber\\
&+n_{\rm T}\frac{\Delta z}{L}\sum^{N}_{j=1}\left(|\xi_{\pm}(z_{j})|^2+|\xi_{\mp}(z_{j})|^2\right)\Bigg]\xi_{\pm}(z).
\end{align}
By integrating these differential equations along a single segment where the slowly-evolving intensities $|\xi_{\pm}|^2$ can be considered as constant in our weak absorption regime and employing the Ansatz~\eqref{eq:HelmholtzAnsatz} one finally arrives to
\begin{align}
\label{eq:FiniteElementAppendix2}
&E_{\pm}(z \pm \Delta z)=exp\Bigg\{i\Bigg[\left(n+i\alpha\frac{c}{\omega}\right)\nonumber\\
&+n_{\rm K}\left(|E_{\pm}(z)|^2+2|E_{\mp}(z)|^2\right)
\nonumber\\
&+n_{\rm T}\frac{\Delta z}{L}\sum^{N}_{j=1}\left(|E_{\pm}(z_{j})|^2+|E_{\mp}(z_{j})|^2\right)\Bigg]\frac{\omega}{c}\Delta z\Bigg\}E_{\pm}(z).
\end{align}
which generalizes Eq.~\eqref{eq:FiniteElement} to generic nonlinearities.

The fields in the different components of the sample are coupled in reciprocal and lossless directional couplers in which the output and input field amplitudes are related by a scattering matrix
\begin{align}
\begin{pmatrix}
    E_{\rm out,1}\\
    E_{\rm out,2}
\end{pmatrix}
=
\begin{pmatrix}
    t_{\rm w,S,m} & ik_{\rm w,S,m} \\
    ik_{\rm w,S,m} & t_{\rm w,S,m}
\end{pmatrix}
\begin{pmatrix}
    E_{\rm in,1}\\
    E_{\rm in,2}
\end{pmatrix}
,
\label{eq:DirectionalCouplers}
\end{align}
where $t_{\rm w,S}$ and $ik_{\rm w,S}$ represent the transmission and coupling amplitudes in the ring-bus and ring-S couplers, respectively. On the other hand, $t_{\rm m}$ and $ik_{\rm m}$ correspond to the transmission and reflection amplitudes at the facets of the bus waveguide, which give rise to Fabry-Pérot oscillations. Note that $t_{\rm w,S,m}$ and $k_{\rm w,S,m}$ are taken as real numbers satisfying $t^2_{\rm w,S,m}+k^2_{\rm w,S,m}=1$. 

Altogether, the set of Eqs.~(\ref{eq:FiniteElementAppendix2}-\ref{eq:DirectionalCouplers}) for all elements of our set-up represent the electric field propagation throughout the sample and can be solved with standard numerical techniques providing a complete and quantitative description of the nonlinear light propagation at the steady-state.

\begin{figure}[t]
    \centering
    \includegraphics[width=0.5\textwidth]{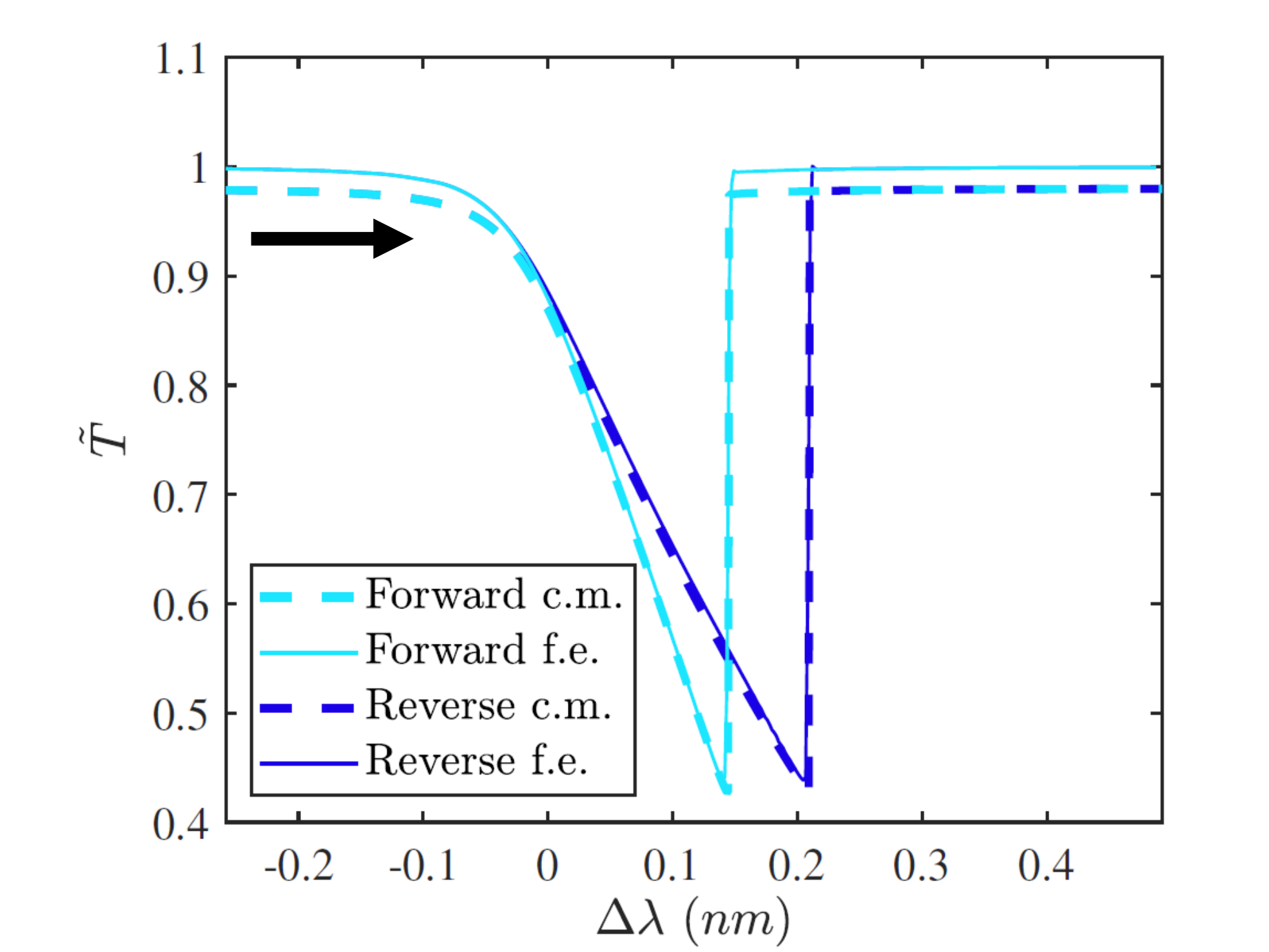}
    \caption{Normalized transmittance $\tilde{T}$ as a function of the relative wavelength shift with respect to the resonance wavelenght ($\Delta\lambda$) in an upwards ramp (indicated by the black arrow) for a TJR in the forward (light blue) and reverse (dark blue) configurations. The dashed curves were calculated by using the coupled-mode theory (c.m.), whereas the solid lines were obtained with the finite-element model (f.e.). The incident power is taken to be $P=0.21$ W. The parameters of the simulations are those of Fig.~\ref{fig:NewFigure3}.}
    \label{fig:ComparisonCM-FE}
\end{figure}
%


\section{Coupled-mode theory and finite-element model in the weak coupling regime}
\label{sec:App_ComparisonCM-FE}

In this Appendix we show that the finite-element simulations recover the coupled-mode theory results in the weak coupling limit ($k_{\rm w,S} \ll 1$) without Fabry-Pérot oscillations ($k_{\rm m}=0$). Fig.~\ref{fig:ComparisonCM-FE} displays the normalized transmittance $\tilde{T}=T/T_{\rm max}$ for a TJR operating in forward and reverse configurations as obtained by using both formalisms. The parameters of the simulations are those employed in Fig.~\ref{fig:NewFigure3} for an incident power $P=0.21$~W. The agreement between both models is best found around resonance where the coupled-mode equations are valid. The discontinuities and the non-reciprocal window are found to lie at the same wavelengths in both simulations. 


\section{Fit parameters}
\label{sec:App_FitParameters}

Table~\ref{tab:FitParameters} summarizes the parameters employed in the fit of the experimental data using the finite-element model derived in Appendix~\ref{sec:App_FE}. A diagram of the simulated device is shown in Fig.~\ref{fig:ModelTaiji}. The TJR is assumed to be at the center of the bus waveguide. Whilst the sample employed in the experiment features an asymmetrical TJR with an optimized shape in order to reduce backscattering of light, to facilitate the calculations we employed a circular TJR with the same ring length.

\begin{table}[h]
\begin{tabular}{|c|c|cccc}
\cline{1-4}
$R_{\rm r}$ & $135.11$ $\mu$m              & \multicolumn{1}{c|}{$n_{\rm L}$} & \multicolumn{1}{c|}{$1.83$}                                   &                      &                      \\ \cline{1-4}
$R_{\rm S}$ & $62.45$ $\mu$m               & \multicolumn{1}{c|}{$n_{\rm K}$} & \multicolumn{1}{c|}{$8 \times 10^{-16}$ cm$^2/$W}             &                      &                      \\ \cline{1-4}
$L_{\rm L}$ & $687.50$ $\mu$m              & \multicolumn{1}{c|}{$n_{\rm T}$} & \multicolumn{1}{c|}{$(8.8 \pm 0.4) \times 10^{-13}$ cm$^2/$W} &                      &                      \\ \cline{1-4}
$L_{\rm R}$ & $687.50$ $\mu$m              & \multicolumn{1}{c|}{$\alpha$}    & \multicolumn{1}{c|}{$0.3454 \times 10^{-4}$ $\mu$m$^{-1}$}    &                      &                      \\ \cline{1-4}
$A$         & $0.66 \times 10^{-8}$ cm$^2$ & \multicolumn{1}{c|}{$k_{\rm w}$} & \multicolumn{1}{c|}{$0.49 \pm 0.02$}                          &                      &                      \\ \cline{1-4}
$N_{\rm r}$ & 8                            & \multicolumn{1}{c|}{$k_{\rm S}$} & \multicolumn{1}{c|}{$0.14 \pm 0.03$}                          &                      &                      \\ \cline{1-4}
$N_{\rm S}$ & 2                            & \multicolumn{1}{c|}{$k_{\rm m}$} & \multicolumn{1}{c|}{$0.24 \pm 0.04$}                          & \multicolumn{1}{l}{} & \multicolumn{1}{l}{} \\ \cline{1-4}
$N_{\rm w}$ & 4                            & \multicolumn{1}{l}{}             & \multicolumn{1}{l}{}                                          & \multicolumn{1}{l}{} & \multicolumn{1}{l}{} \\ \cline{1-2}
\end{tabular}
\caption{Fit parameters: ring radius $R_{\rm r}$, S waveguide radius $R_{\rm S}$, bus waveguide length to the left $L_{\rm L}$ and right $L_{\rm R}$ of the TJR, waveguide cross section $A$, number of segments in which the ring, S and bus waveguides are divided $N_{\rm r,S,w}$, linear refractive index $n_{\rm L}$, nonlinear Kerr refractive coefficient $n_{\rm K}$, nonlinear thermal refractive coefficient $n_{\rm T}$, absorption losses $\alpha$, coupling parameters $k_{\rm w,S}$ for the ring-bus waveguide and ring-S waveguide couplers, and reflection amplitude $k_{\rm m}$ at the bus waveguide facets.}
\label{tab:FitParameters}
\end{table}
%


\section{Resonance shift for equivalent thermo-optic and Kerr nonlinearities}
\label{sec:App_ShiftThermalKerr}

\begin{figure}[h]
    \centering
    \includegraphics[width=0.5\textwidth]{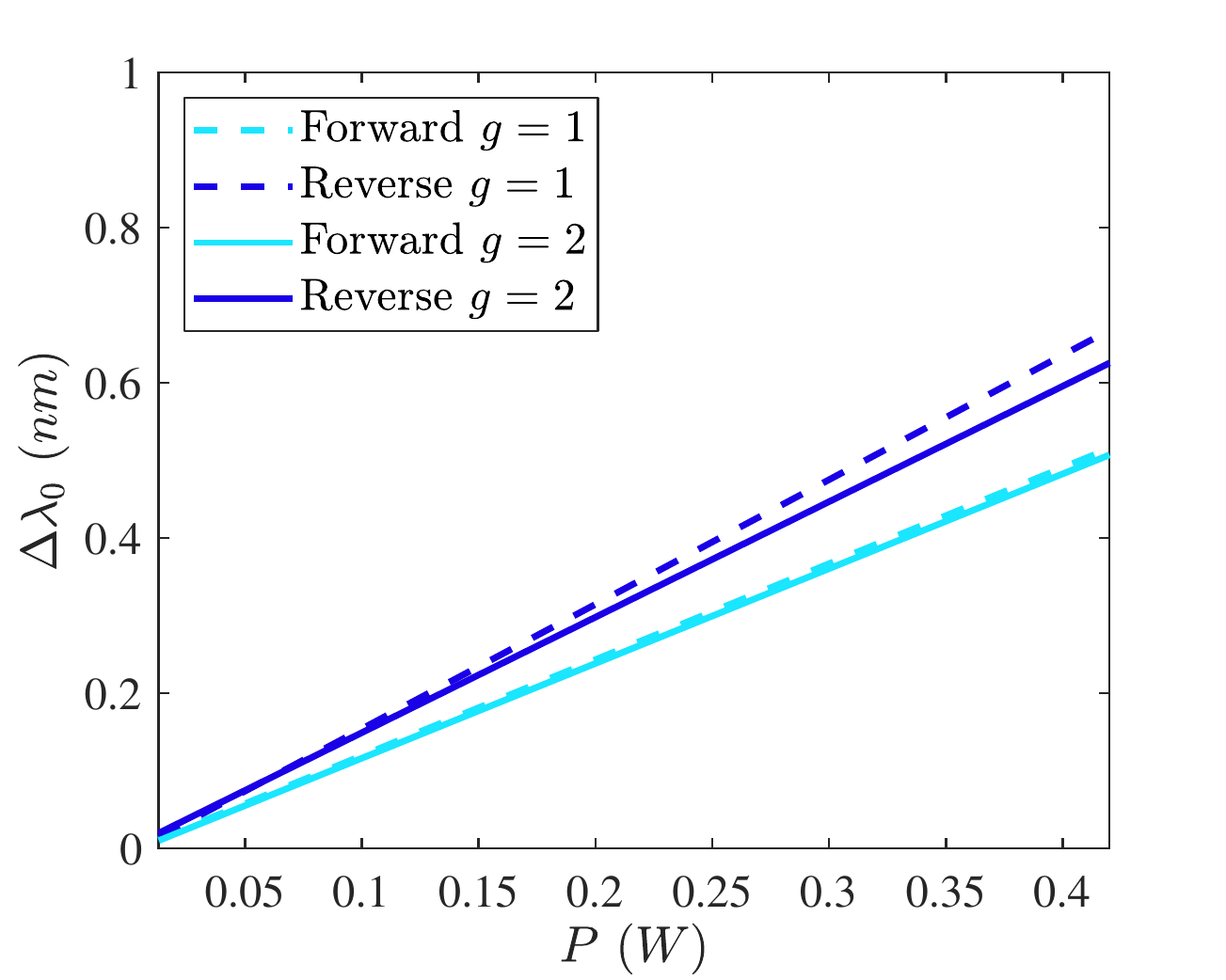}
    \caption{Shift of the resonance wavelength $\Delta\lambda_{0}$ w.r.t. its linear value $\lambda=1545.76$~nm as a function of the input power $P$. All curves are linear fits of the numerically calculated data using the finite-element model. Dashed lines are the curves displayed in Fig.~\ref{fig:ReciprocityBreaking}b  for the experimental case of a thermo-optic nonlinearity with $n_{\rm T}=8.8\times 10^{-13}$~cm$^2/$W. Solid lines are obtained by employing a purely Kerr nonlinearity ($n_{\rm T}=0$) whose strength has been artificially increased to $n_{\rm K}=8.8\times 10^{-13}$~cm$^2/$W to  match the strength of the thermo-optic nonlinearity of the experiment. 
    Light (dark) blue lines correspond to the forward (reverse) configuration.}
    \label{fig:ResonanceShiftThermalKerr}
\end{figure}

Here we compare the resonance shift $\Delta\lambda_{0}$ produced by the thermo-optic nonlinearity displayed by our samples with the one that would exhibit a (fictitious) TJR featuring a Kerr nonlinear parameter of the same magnitude $n_{\rm K}=8.8\times 10^{-13}$~cm$^2/$W and a negligible $n_{\rm T}\simeq 0$. In all calculations, $\Delta\lambda_{0}$ is measured w.r.t. the linear position of the resonance at $\lambda=1545.76$~nm.
Fig.~\ref{fig:ResonanceShiftThermalKerr} shows linear fits of the numerical results of the finite-element model in each case, which slightly deviate from the linear behaviour due to Fabry-Pérot oscillations: quite unexpectedly, for a given value of the input power the Kerr nonlinearity with $g=2$ gives a slightly smaller nonlinear shift $\Delta\lambda_0$ than the thermo-optic nonlinearity ($g=1$). 

These numerical results can be physically understood using the coupled-mode equations~(\ref{eq:CoupledMode1}-\ref{eq:CoupledMode2}). In the forward configuration light circulates in the CCW mode only and no significant difference between both kinds of nonlinearity arises. 
In the reverse configuration, one could have expected that the presence of the $g=2$ factor in the Kerr nonlinear term of Eq.~\eqref{eq:CoupledMode2} would give a larger nonlinear shift $\Delta\lambda_{0}$ compared to the thermo-optical case. The complete calculation displayed here shows that this is not the case, since the very presence of this factor $g=2$ 
quickly pushes the CCW mode out of resonance from the CW one and the pump laser. This results in a smaller value of the intensity in the CCW mode, which may well overcompensate the factor 2.
As a result, the net effective nonlinear shift $\Delta\lambda_{0}$ turns out to be a bit smaller for a Kerr nonlinearity than for a thermal one at the same input power.

%


\bibliography{NonlinearTaiji}

\end{document}